\documentclass{ifacconf}

\usepackage{graphicx}      
\usepackage{natbib}        
\usepackage{amsmath}       
\usepackage{amssymb}       
\usepackage{dsfont,stmaryrd}
\usepackage{algorithm}
\floatname{algorithm}{Alg.}
\usepackage{algpseudocode}
\input{main.sty}
\usepackage{multirow} 
\usepackage{array}    
\usepackage{graphicx}
\usepackage{lipsum}
\usepackage{todonotes}
\usepackage{enumerate}
\usepackage{booktabs}
\usepackage[absolute,overlay]{textpos} 
\begin{document}
\begin{textblock*}{15cm}(2cm,2cm) 
This work has been submitted to IFAC for possible publication
\end{textblock*}
\begin{frontmatter}

\title{Data-Selective Online Battery Identification Using Extended Time Regular Expressions} 

\thanks[footnoteinfo]{This work was supported by the Villum Foundation for the Smart Battery project (Project No. 222860).}

\author[First]{Nicolai A. Weinreich} 
\author[Second]{Marco Muñiz} 
\author[Second]{Marius Mikučionis}
\author[Second]{Kim G. Larsen}
\author[First]{Remus Teodorescu}

\address[First]{Department of Energy, Aalborg University, Aalborg, Denmark, (email: \{nawe, ret\}@energy.aau.dk)}
\address[Second]{Department of Computer Science, Aalborg University, Aalborg, Denmark, (email: \{muniz, marius, kgl\}@cs.aau.dk)}

\begin{abstract}                
In this paper, we propose a data-efficient online battery identification method which targets highly informative
battery cell data segments based on the driving pattern of the vehicle.
We consider the case of a vehicle driving on/off a motorway and construct an Extended Time Regular Expression (ETRE) to detect data segments fitting these driving patterns.
Simulation results indicate that by only using up to 10.71\% of the data on average, the proposed method provides a low-bias and low-variance estimator under non-negligible current and voltage noise compared to other conventional estimation algorithms.

\end{abstract}


\end{frontmatter}

\section{Introduction}

Accurate battery modeling and parameter identification is critical for effective cell monitoring and management. 
This is especially important in Electric Vehicle (EV) applications due to concerns about safety, available driving range and longevity of the battery pack.
Online identification methods are crucial for this due to their application on computationally simple models and ability to adapt these models to different operating conditions such as State of Charge (SOC), temperature, and aging based on real-world cell current/voltage data.
However, unlike laboratory-generated battery data, real-world data introduce several challenges such as noise corruption and periods of low informativity.
The development of online identification methods which are robust towards these challenges and computationally feasible on on-board hardware is highly important.

Conventional online identification methods involve modeling a battery cell as a linear system and iteratively estimating the system parameters using a variant of Recursive Least Squares (RLS). 
While RLS is computationally simple, it has two notable challenges: data saturation, i.e. recent parameters are estimated based on old data, and estimation bias under input noise.
To minimize data saturation, a forgetting factor is commonly introduced to weigh newer data higher (\cite{lao_novel_2018,zhao_lithium_2024}). To handle noise corruption, noise-correcting methods have been introduced, as in \cite{trongnukul_robust_2025}. 
Other works have proposed using Total Least Squares (TLS) instead of RLS, due to it naturally accounting for input noise (\cite{wei_online_2018,du_online_2023}), albeit with a greater computational cost over time.
In addition, while the TLS produces less biased estimates, it is still susceptible to data with low informational content which may increase the variance, and therefore the reliability, of the estimate.

Recently, data-selective methods have been proposed to quantify the informativity of the measured data, using Fisher information, and only update model parameters 
when informative data is present (\cite{du_information_2022,fogelquist_data_2023,cai_novel_2024}). However, the Fisher information measures in present works are mostly 
defined in a least-squares setting under the assumption of negligible noise on the input, the inclusion of which makes direct calculation of the Fisher information impractical in an online setting. 
In \cite{weinreich_data_2025}, the Fisher information under the TLS framework was used in a simulated environment to train a neural network in predicting when informative cell data is present using EV acceleration data as a proxy.
The main benefit of predicting informativity from proxy driving data such as speed and acceleration is that the decision of using measured data to update a parameter is off-loaded from multiple 
low-complexity hardware, such as cell measurement boards, to a centralized system with more computational power. 

In this paper, we aim to target segments of cell data with a high informational content to be used for online estimation of battery parameters.
Unlike data-selective methods based on an information measure, we target data segments based on the driving conditions of the EV and a pre-defined notion of which types of driving yield more informative data.
Specifically, we consider the typical case of driving on and off a motorway. 
We expect the transition between highway and motorway driving to include high acceleration/deceleration segments resulting in higher current excitations and thus more informative cell data. 
By formally defining an expression of these transition segments, it is possible target them during EV operation.

Regular Expressions (REs)~(\cite{kleene1956regex}) is a
fundamental concept in Computer Science with wide industrial
application. One of the main applications is in 
pattern matching in which a RE (a pattern) is matched for occurrences in a large text. 
Examples include matching all emails, telephone numbers, etc.\ in a document. 
The work in~\cite{AsarinKleeneTheorem} extends RE with timing constraints yielding a timed RE
(TRE) and in~\cite{etrepaper} an extended TRE (ETRE) is formulated with arbitrary
functions and real valued inputs, making it possible to perform
matching of complex timed patterns in time series data.
A specially crafted ETRE matching entering and leaving a motorway allows us to target the more informative speed segments, to skip low information updates, and to obtain better parameter estimates with less data.

The effectiveness of the method is demonstrated in a simulated environment of EVs driving on a model of a real motorway in North Jutland, Denmark.

\section{Online Battery Modeling and Estimation}

\begin{figure}[t]
\begin{center}
    \includegraphics[width=.8\columnwidth]{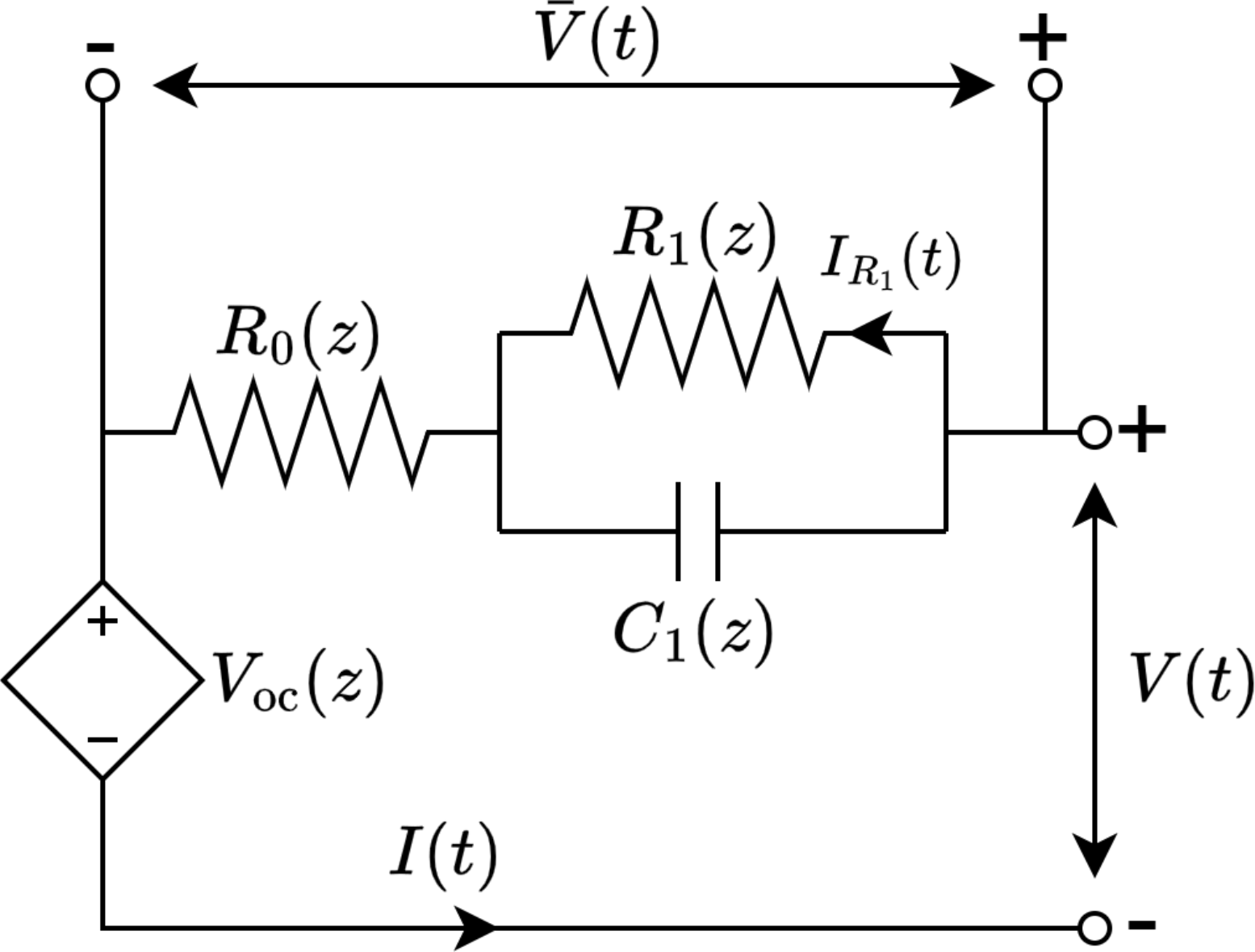}
    \caption{Electrical battery model used in this paper.}
    \label{fig: ecm}
\end{center}
\end{figure}
In this section, the battery model used in the paper is defined along with the RLS and TLS algorithm.
A typical 1RC Electrical Circuit Model (ECM) of a battery is shown in Fig. \ref{fig: ecm}. The circuit consists of a voltage source representing the Open-Circuit Voltage (OCV) $V_{\mathrm{oc}} : \mathbb{R} \to \mathbb{R}$, a series resistance $R_0 : \mathbb{R} \to \mathbb{R}$, and an RC pair of resistance $R_1 : \mathbb{R} \to \mathbb{R}$ and capacitance $C_1 : \mathbb{R} \to \mathbb{R}$. 
Note that the OCV and the electrical parameters depend on the SOC, $z\in \mathbb{R}$.

The terminal voltage of the cell is modeled as
\begin{align}
    V(t) &= V_{\mathrm{oc}}\big(z(t)\big) + R_0\big(z(t)\big)I(t) + R_1\big(z(t)\big)I_{R_1}(t)\\
    &= V_{\mathrm{oc}}\big(z(t)\big) + \bar{V}(t) \label{eq: an_overpotential_appeared}
\end{align}
where $\bar{V}(t)$ is the overpotential over the series resistance and RC pair. By employing Laplace transformation to the ECM, the overpotential is modeled as a 3$^\mathrm{rd}$ order autoregressive with exogeneous input (ARX) model
\begin{equation}\label{eq: ARX}
    \bar{V}_{k} = \theta_{1,k}\bar{V}_{k-1} + \theta_{2,k}I_k + \theta_{3,k} I_{k-1} = \boldsymbol{\theta}^T_k\boldsymbol{\phi}_k
\end{equation}
where $\boldsymbol{\theta}_k = [\theta_{1,k}, \theta_{2,k}, \theta_{3,k}]^T$, $\boldsymbol{\phi}_k = [\bar{V}_{k-1}, I_k, I_{k-1}]^T$, and $k\in \mathbb{Z}$ is an index variable. The ARX parameters can be expressed by the ECM parameters as
\begin{align}
    \theta_{1,k} &= \frac{2R_1C_1-T}{2R_1C_1+T} \label{eq: theta_1}\\
    \theta_{2,k} &= R_0 + \frac{R_1T}{2R_1C_1 + T}\\
    \theta_{3,k} &= \frac{(R_0+R_1)T-2R_0R_1C_1}{2R_1C_1+T}\label{eq: theta_3}
\end{align}
where $T$ is the sampling interval and the dependence of SOC is omitted for brevity. 
The goal is to track $\boldsymbol{\theta}_k$ over time based on measured input current and terminal voltage.

Using the expression in \eqref{eq: ARX}, $\boldsymbol{\theta}_k$ is estimated using a least-squares approach. 
For online estimation, RLS is typically used with a forgetting factor $0 \ll \lambda <1 $. 
The RLS algorithm is summarized in Algorithm \ref{alg: RLS}. While RLS is computationally simple, it is well known to produce biased estimates under noise-corrupted input. Alternatively, the TLS estimation method can be used. 
For a small enough number of samples, $L$, it can be assumed that the ARX parameters are static i.e. $\boldsymbol{\theta}_k = \boldsymbol{\theta}_{k-1} = \cdots = \boldsymbol{\theta}_{k-L+1}$. Given a segment of $L$ data points, we construct the system

\begin{equation}\label{eq: TLS_system}
     \underbrace{\begin{bmatrix}
        \bar{V}_k\\
        \bar{V}_{k-1}\\
        \vdots\\
        \bar{V}_{k-L+2}
     \end{bmatrix}}_{\boldsymbol{y}_k} = \underbrace{\begin{bmatrix}
        \bar{V}_{k-1} & I_{k} & I_{k-1} \\
        \bar{V}_{k-2} & I_{k-1} & I_{k-2}\\
        &\vdots&\\
        \bar{V}_{k-L+1} & I_{k-L+2} & I_{k-L+1}
     \end{bmatrix}}_{\boldsymbol{X}_k}\begin{bmatrix}
        \theta_{1,k}\\
        \theta_{2,k}\\
        \theta_{3,k}
     \end{bmatrix},\quad \forall k
\end{equation}
where $\boldsymbol{y}_k \in \mathbb{R}^{L-1}$ and $\boldsymbol{X}_k \in \mathbb{R}^{(L-1) \times 3}$. The TLS estimate of $\boldsymbol{\theta}_k$ is then given by Algorithm \ref{alg: TLS}. 
The TLS estimate is less biased under input noise; however, it requires computing a Singular Value Decomposition (SVD) which in our case has a computational complexity of $\mathcal{O}(L)$. Additionally, for small $L$, the TLS estimate may have a high variance depending on the informativity of the data in the segment.

\begin{algorithm}[t]
\caption{$[\hat{\boldsymbol{\theta}}_{k},\mathbf{P}_{k}]$ = RLS$[\hat{\boldsymbol{\theta}}_{k-1},\mathbf{P}_{k-1}, \bar{V}_{k},\boldsymbol{\phi}_k, \lambda]$}\label{alg: RLS}
\begin{algorithmic}[1]
\State Calculate gain: $\mathbf{L}_k = \mathbf{P}_{k-1}\boldsymbol{\phi}_k\big/(\lambda+\boldsymbol{\phi}_k^T\mathbf{P}_{k-1}\boldsymbol{\phi}_k)$
\State Update parameter: $\hat{\boldsymbol{\theta}}_k = \hat{\boldsymbol{\theta}}_{k-1} + \mathbf{L}_k(\bar{V}_{k}-\hat{\boldsymbol{\theta}}_{k-1}^T\boldsymbol{\phi}_k)$
\State Update covariance: $\mathbf{P}_k = \frac{1}{\lambda}(\mathbf{I}-\mathbf{L}_k\boldsymbol{\phi}_k^T)\mathbf{P}_{k-1}$
\end{algorithmic}
\label{alg: RLS}
\end{algorithm}

\begin{algorithm}[t]
\caption{$\hat{\boldsymbol{\theta}}_k$ = TLS$[\boldsymbol{y}_k, \boldsymbol{X}_k]$}\label{alg: TLS}
\begin{algorithmic}[1]
\State Construct augmented matrix: $\boldsymbol{H} =\begin{bmatrix}
    \boldsymbol{X}_k&\boldsymbol{y}_k
\end{bmatrix} $
\State Perform SVD: $\boldsymbol{H}=\boldsymbol{U}\boldsymbol{\Lambda}\boldsymbol{V}=\boldsymbol{U}\boldsymbol{\Lambda}\begin{bmatrix}
        \mathbf{V}_{pp}&\boldsymbol{v}_{pq}\\
        \boldsymbol{v}_{qp}&v_{qq}
    \end{bmatrix}$
\State Compute parameter: $\hat{\boldsymbol{\theta}}_{k} = -\boldsymbol{v}_{pq}/v_{qq}$
\end{algorithmic}
\label{alg: TLS}
\end{algorithm}

\section{Extended Time Regular Expression of Highway/Motorway Transitions}

\begin{figure}[t]
    \begin{center}
        \includegraphics[width=\columnwidth]{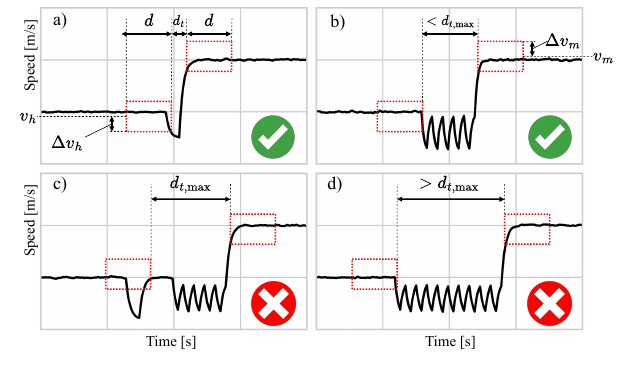}
        \caption{Examples of the ETRE, $\trelow_{hm}$, being applied to speed profiles
          of an EV going from the highway to the motorway.  In a) and
          b), the profiles match the expression since the speed values
          are contained in the defined tubes and the duration between
          the tubes is less than $d_{t,\mathrm{max}}$. In c), not all speed
          values are contained in the first tube and in d), the
          duration between the tubes is too long, resulting in no
          matches.}
      \end{center}
      \label{fig:examplespeed}
\end{figure}

In this section, the ETRE of an EV driving on/off the motorway is defined.
Following the grammar defined in \cite{etrepaper}, we define the following expressions
\begin{align}
    \trelow_h &= \langle({v_h}_{\Delta v_h})^+\rangle_{[d,d]}\label{eq: etre_highway}\\
    \trelow_m &= \langle({v_m}_{\Delta v_m})^+\rangle_{[d,d]}\\
    \trelow_t &= \langle \Sigma^*\rangle_{[0,d_{t,\mathrm{max}}]}\label{eq: etre_transition}
\end{align}

The expression $\trelow_h$ describes sequences where the speed of a vehicle is $v_h\pm \Delta v_h$ for a
duration between $d$ to $d$ seconds which practically means at least $d$ seconds.
This expression is used to capture speed data where the EV is driving continously on the highway.

The expression $\trelow_m$ describes sequences where the speed of a vehicle is $v_m\pm \Delta v_m$ for at least $d$ seconds. 
This expression is used to capture speed data where the EV is driving continously on the motorway.

The expression $\trelow_t$ describes sequences of all possible speeds for a duration between 0 and $d_{t,\mathrm{max}}$ seconds and is used to capture the transition between the highway and motorway.

By combining the expressions in \eqref{eq: etre_highway}-\eqref{eq: etre_transition}, we construct an ETRE describing the EV driving on and off a motorway.
For instance, the ETRE $\tre_{hm} = \trelow_h \cdot \trelow_t \cdot \trelow_m$ describes the EV driving from the highway and onto the motorway. 
Visually, $\tre_{hm}$ can be represented as 2-dimensional tubes which both have to fully contain parts of the speed time series data while being less than $d_{t,\mathrm{max}}$ seconds apart. 
Figure \ref{fig:examplespeed} four examples of a time series signal matchhing and not matching the ETRE.

In the remainder of this paper, we define the ETRE as 
\begin{equation}\label{eq: etre}
    \tre = \trelow_h \cdot \trelow_t \cdot \trelow_m \mid \trelow_m \cdot \trelow_t \cdot \trelow_h = \tre_{hm} \mid \tre_{mh}
\end{equation}
which describes the EV driving from the highway onto the motorway or vice versa. 
We define $\llbracket\tre\rrbracket$ as the set of all possible time series signals which match $\tre$.

\section{Data-Selective TLS using ETRE}

\begin{figure}[t]
\begin{center}
    \includegraphics[width=\columnwidth]{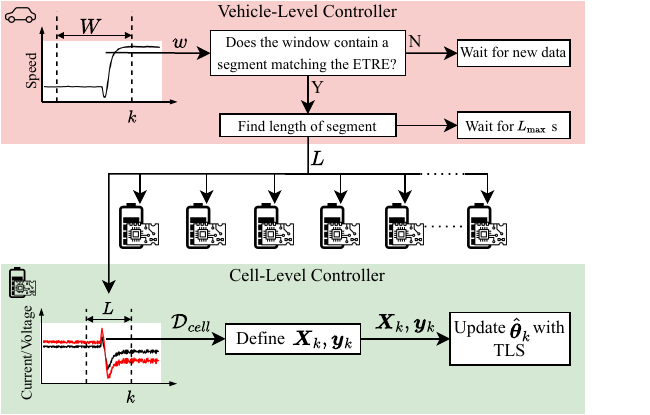}
    \caption{Flow diagram of the proposed methodology in this paper.}
    \label{fig: framework}
\end{center}
\end{figure}

In this section, we propose the online estimation method, named Data-Selective TLS (DS-TLS). 
The key idea of the method is to update $\hat{\boldsymbol{\theta}}$ only informative cell measurements have been generated, based on the speed profile of the EV.
The proposed method is shown in Fig. \ref{fig: framework}. First, an observation window $w_k = \{v_{k-W+1},\dots,v_k\}$ of length $W$ is used to monitor a stream of time series speed data.
At time $k$, the data in the observation window is passed to the timed pattern matching algorithm described in \cite{etrepaper} using the ETRE defined 
in \eqref{eq: etre} to find speed segments $s \subset w$ such that $s \in \llbracket\tre\rrbracket$. 
If no segments $s \in \llbracket\tre\rrbracket$ are detected, then the algorithm waits for the next sample of speed data, and we keep the previous estimate i.e. $\hat{\boldsymbol{\theta}}_k = \hat{\boldsymbol{\theta}}_{k-1}$. 
If a segment is detected, then its length $L = |s|$, where $|\cdot|$ denotes cardinality, is transmitted to the cell boards which in turn collect their respective cell data
\begin{equation}
    \mathcal{D}_{cell} = \{I_{k-L+1}, \dots, I_{k},V_{k-L+1},\dots,V_{k}\}.
\end{equation}
We note that the segment length $L$ is not necessarily a fixed number due to the expression in \eqref{eq: etre_transition}. In fact, $L_{\mathrm{min}} \leq L \leq L_{\mathrm{max}}$ where $L_{\mathrm{min}} = 2d$ and $L_{\mathrm{max}}=2d+d_{t,\mathrm{max}}$.

After the segment of cell data has been collected, the overpotentials $\bar{V}_{k-L+1},\dots,\bar{V}_{k}$ are estimated using \eqref{eq: an_overpotential_appeared}.
Note that an accurate measurement of the overpotential requires an accurate measurement of, $z$, and the mapping $V_{\mathrm{oc}}(\cdot)$. 
For the purpose of this paper, these are assumed to be known; however, in real applications they have to be replaced by estimated values.
The observation matrix, $\boldsymbol{X}_k$, and output vector, $\boldsymbol{y}_k$, are then defined as in \eqref{eq: TLS_system} and TLS is used to estimate $\hat{\boldsymbol{\theta}}_k$.
After a speed segment matching $\tre$ has been found, the data selection algorithm waits for $L_{\mathrm{max}}$ samples before looking for new matches. 
This is done to prevent overlapping segments of data being used in the estimation process and save computational resources.

To mitigate the effect of poor parameter initialization, the DS-TLS algorithm always updates $\hat{\boldsymbol{\theta}}$ using the first $L_{\mathrm{max}}$ cell measurements of the trip.

\section{Experimental Setup}

\begin{figure}[t]
\begin{center}
    \includegraphics[width=.8\columnwidth]{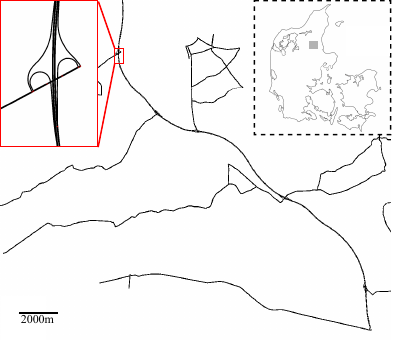}
    \caption{Section of the road network map used.}
    \label{fig: E45_section}
\end{center}
\end{figure}

A road network modeling a $134$ km segment of the E45 motorway in North Jutland, Denmark was constructed in the urban mobility simulator SUMO (see \cite{lopez_microscopic_2018}). 
The network includes highways leading to/from the motorway.
A section of the road network is shown in in Fig. \ref{fig: E45_section}.
Using this network, speed profiles of EVs repeatedly driving to and from the motorway were generated.
Each EV was modeled using a 120s4p battery pack with a nominal energy capacity of 88 kWh. 
The cells in the battery pack were simulated from $95\%$ to $5\%$ SOC using a 1RC ECM of a 50Ah Li-ion prismatic cell as specified in \cite{zheng_online_2024}. 
Each cell is assumed to be identical with a constant temperature of 25 degrees Celsius.
The EV, battery pack, and battery cells were simulated using the python package TrackSim (see \cite{weinreich_tracksim_2025}). True ARX parameters were found using \eqref{eq: theta_1}-\eqref{eq: theta_3}. 
The true overpotential was then defined by \eqref{eq: ARX} and used to define the true terminal voltage via \eqref{eq: an_overpotential_appeared}.
The settings for the ETRE expressions are listed in Table \ref{tab: parameter_settings}. These values were found by analyzing the highway and motorway sections of the generated speed profiles. 
Settings for the RLS algorithm are given in Table \ref{tab: parameter_settings}. For TLS, consecutive data segments of length $L_{\mathrm{TLS}}$ were used to update $\hat{\boldsymbol{\theta}}_k$. 
To enable fair comparison between TLS and DS-TLS, for each trip we define
\begin{equation}
    L_{\mathrm{TLS}} = \left\lfloor\frac{1}{|\mathcal{S}|}\sum_{s_i\in\mathcal{S}}L_i\right\rceil
\end{equation}
where $\mathcal{S}$ is the set of all segments selected by DS-TLS, $L_i=|s_i|$, and $\lfloor\cdot\rceil$ denotes rounding to the nearest integer.

\begin{figure}[ht]
\begin{center}
    \includegraphics[width=\columnwidth]{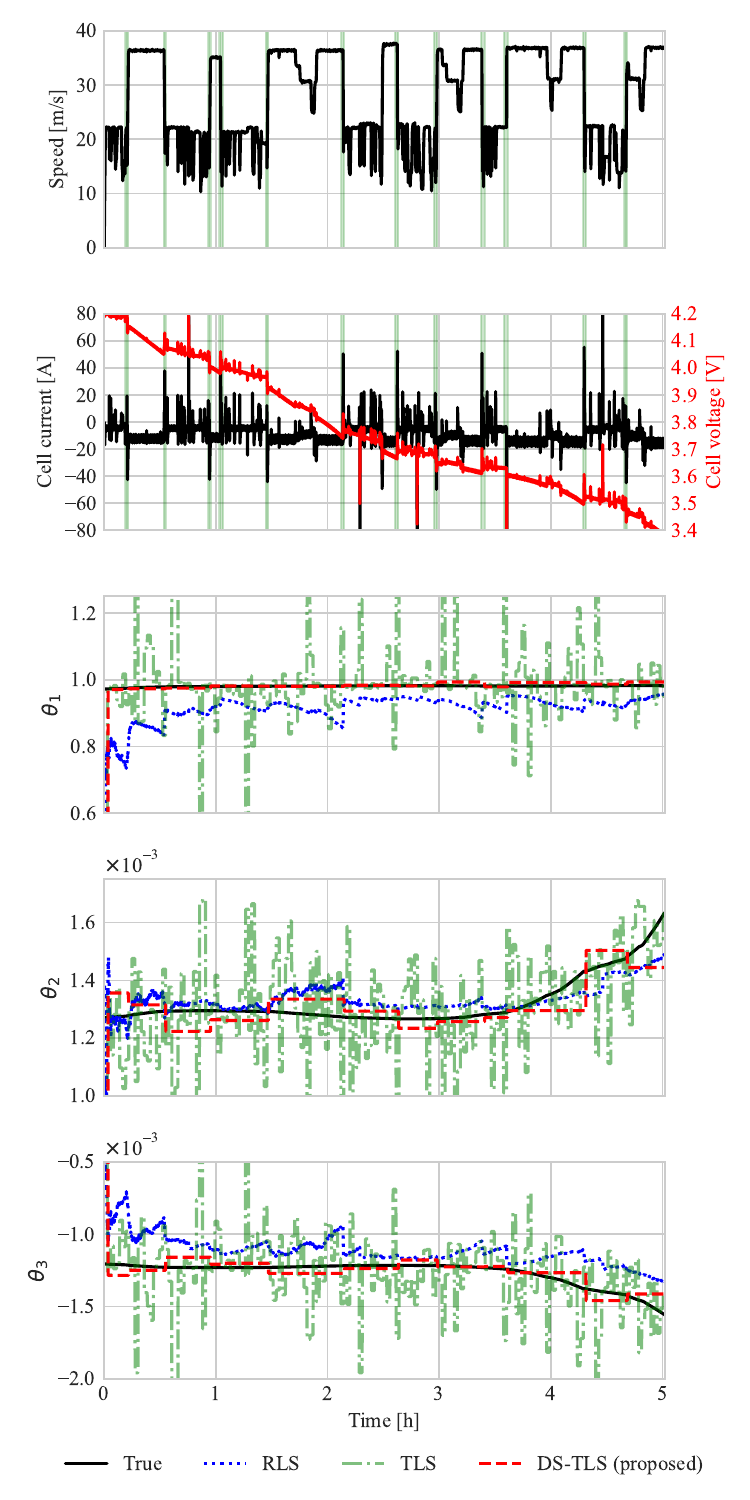}
    \caption{Estimation results on example trip with $d=30$s. 
    Selected segments are highlighted in green in the speed and current/voltage profiles.}
    \label{fig: example_trip}
\end{center}
\end{figure}

\begin{table}[t]
\begin{center}
\caption{Experiment settings.}\label{tab: parameter_settings}
\renewcommand{\arraystretch}{1} 
\begin{tabular}{ccc}
\toprule
 & {\bf Parameter} & {\bf Setting}\\\midrule
\multirow{6}{*}{\rotatebox{90}{ETRE}} & $v_{h}$ & 20 m/s (72 km/h)\\
&$\Delta v_{h}$ & 5 m/s (18 km/h)\\
&$v_{m}$ & 34 m/s (122.4 km/h)\\
&$\Delta v_{m}$ & 10 m/s (36 km/h)\\
&$d_{t,\mathrm{max}}$ & 60s\\
&$d$ & $\{10, 30, 60, 120, 280, 240, 300\}$s\\\midrule
\multirow{3}{*}{\rotatebox{90}{RLS}}&$\hat{\boldsymbol{\theta}}_0$ & $[0, 0, 0]^T$\\
&$\mathbf{P}_0$ & $\text{diag}(10^6, 10^6, 10^6)$\\
&$\lambda$ & 0.999\\
\bottomrule
\end{tabular}
\end{center}
\end{table}

\begin{table*}[ht!]
\begin{center}
\caption{Simulation results for all trips under different choices of $d$. 
The values are averaged over all trips $\pm$ 1 standard deviation. 
Minimum values for DS-TLS are highlighted in bold.}\label{tab: results}
\renewcommand{\arraystretch}{1} 
\begin{tabular}{ccccccc}
\toprule
Est. Alg. & $d$ [s] & $\theta_1$ MAPE [\%]& $\theta_2$ MAPE [\%]& $\theta_3$ MAPE [\%] & $V$ RMSE [mV] & Data usage [\%]\\\midrule
RLS & - & $8.94 \pm 1.24$ & $3.73 \pm 0.68$ & $14.44 \pm 2.00$ & $1.88 \pm 0.01$ & -\\\midrule
TLS & 10 & $23.21 \pm 33.31$ & $22.94 \pm 20.61$ & $42.59 \pm 48.58$ & $13.75 \pm 47.13$ & -\\
DS-TLS & 10 & $1.82 \pm 2.42$ & $6.53 \pm 2.03$ & $7.95 \pm 3.85$ & $2.26 \pm 1.46$ & $\mathbf{4.80 \pm 0.82}$\\\midrule
TLS & 30 & $8.17 \pm 10.98$ & $11.23 \pm 8.30$ & $18.18 \pm 17.99$ & $4.10 \pm 19.23$ & -\\
DS-TLS & 30 & $0.62 \pm 0.26$ & $5.42 \pm 0.70$ & $6.09 \pm 0.83$ & $2.00 \pm 0.05$ & $7.70 \pm 1.17$\\\midrule
TLS & 60 & $7.27 \pm 91.65$ & $8.53 \pm 32.99$ & $15.99 \pm 150.54$ & $4.75 \pm 60.06$ & -\\
DS-TLS & 60 & $0.41 \pm 0.11$ & $5.01 \pm 0.66$ & $5.48 \pm 0.73$ & $\mathbf{2.00 \pm 0.02}$ & $10.71 \pm 1.81$\\\midrule
TLS & 120 & $2.55 \pm 0.79$ & $5.78 \pm 0.56$ & $7.86 \pm 1.16$ & $2.11 \pm 0.36$ & -\\
DS-TLS & 120 & $\mathbf{0.34 \pm 0.13}$ & $5.04 \pm 0.83$ & $5.41 \pm 0.91$ & $2.00 \pm 0.03$ & $9.59 \pm 3.09$\\\midrule
TLS & 180 & $1.56 \pm 0.28$ & $4.39 \pm 0.27$ & $5.65 \pm 0.45$ & $2.02 \pm 0.05$ & -\\
DS-TLS & 180 & $0.34 \pm 0.14$ & $\mathbf{4.97 \pm 0.88}$ & $\mathbf{5.28 \pm 0.96}$ & $2.01 \pm 0.04$ & $9.74 \pm 3.72$\\\midrule
TLS & 240 & $1.16 \pm 0.23$ & $3.71 \pm 0.22$ & $4.61 \pm 0.36$ & $2.00 \pm 0.02$ & -\\
DS-TLS & 240 & $0.39 \pm 0.17$ & $5.18 \pm 0.96$ & $5.51 \pm 1.07$ & $2.02 \pm 0.07$ & $7.02 \pm 3.52$\\\midrule
TLS & 300 & $0.79 \pm 0.17$ & $3.13 \pm 0.20$ & $3.68 \pm 0.29$ & $1.99 \pm 0.01$ & -\\
DS-TLS & 300 & $0.38 \pm 0.15$ & $5.08 \pm 0.91$ & $5.38 \pm 0.99$ & $2.02 \pm 0.08$ & $7.14 \pm 3.76$\\\bottomrule
\end{tabular}
\end{center}
\end{table*}

To assess the performance of the estimation algorithms, three metrics are used. First, the Root Mean Squared Error (RMSE) of the predicted terminal voltage is defined as
\begin{equation}\label{eq: RMSE}
    \mathrm{RMSE} = \sqrt{\frac{1}{K-L_{\mathrm{max}}} \sum_{k=L_{\mathrm{max}}+1}^K\left(V_{t,k}-\hat{V}_{t,k}\right)^2}
\end{equation}
where $\hat{V}_{t,k}$ is the predicted terminal voltage at time $k$ and $K$ is the length of the trip. 
Note errors are only measured for $k>L_{\mathrm{max}}$ in order to ignore initialization errors.
To assess the estimation performance on the ARX parameters, we use the Mean Absolute Percentage Error (MAPE) given by
\begin{equation}\label{eq: MAPE}
    \mathrm{MAPE}_i = \frac{100}{K-L_{\mathrm{max}}}\sum_{k=L_{\mathrm{max}}+1}^{K}\left|\frac{\theta_{i,k}-\hat{\theta}_{i,k}}{\theta_{i,k}}\right|, \quad i=1,2,3.
\end{equation}
\newpage
For DS-TLS, the data usage is defined as
\begin{equation}\label{eq: data_usage}
    \mathrm{DU} = \frac{100}{K}\sum_{s_i\in \mathcal{S}}L_i.
\end{equation}
To represent an environment with non-negligible noise, zero-mean Gaussian noise was added to the true 
current and voltage with standard deviations of $\sigma_I = 0.02$ A and $\sigma_V = 0.002$ V. 
To make sure that the estimation results are not sensitive to the specific realization of the noise, the 
estimation process for each trip is repeated $10$ times in a Monte Carlo fashion with different noise realizations.
The metrics defined in \eqref{eq: RMSE} and \eqref{eq: MAPE} are then averaged across these $10$ trials for each trip.

\section{Results}

Figure \ref{fig: example_trip} shows the estimation results on a simulated trip with $d=30$s. 
As seen in the speed profile, the defined ETRE is able to capture most of the highway/motorway transitions.
For $\theta_1$ and $\theta_3$, the RLS estimates exhibit clear biases which increase under periods of low informativity.
The TLS estimates are less biased but vary a lot depending on the segment used. 
By only updating $\hat{\boldsymbol{\theta}}_k$ with the selected segments, both the 
bias and the variance of the estimate are significantly reduced, leading to overall better tracking of $\boldsymbol{\theta}_k$.

The estimation results over 917 simulated trips are shown in Table \ref{tab: results}. 
For $d>30$s, similar estimation errors on $\theta_1$ and $\theta_3$ are achieved 
with average MAPEs of $0.34\%-0.41\%$ and $5.28\%-5.51\%$ respectively 
while only using $7.02\%-10.71\%$ of the data during the trip. Reasonable tracking of $\theta_2$ 
can also be observed for DS-TLS with an average MAPE of $4.97\%-5.18\%$ 
although RLS still outperforms DS-TLS in this regard.

While the DS-TLS algorithm performs best for $d>30$s, recall that a higher $d$ means longer data segments 
and therefore bigger $\boldsymbol{H}$ in the SVD of Alg. \ref{alg: TLS}, leading to higher computational costs.
Subject to hardware limitations, it may suffice to choose $d=30$s or even $d=10$s which will significantly reduce 
the computational burden of the SVD while still yielding reasonable estimation results.
We also note that for high choice of $d$, the resulting ETRE will become more restrictive.
Thus, a higher number of transitions will be ``missed" in the selection process, leading to fewer updates overall.

For $d<120$s, note that the non-selective TLS estimates have high standard deviations.
This can be attributed to the high variance of the TLS estimator due to higher likelihood of encountering non-informative segments 
which yield bad estimates, as is also observed in Fig. \ref{fig: example_trip}.
To compensate for this, the TLS algorithm needs longer segments in the estimation process resulting in more computations.
By only using informative segments, this high variance is effectively reduced, thus enabling the use of TLS on small segments 
and reducing the amount of computations required.

\section{Conclusion}

Data-selective estimation methods have shown a large potential of enhanced online estimation performance by only focusing on informative data, thus ensuring greater data- and energy efficiency.
By moving the selection decision from each individual cell to the vehicle-level and targeting segments where the driving pattern induces informativity in the battery cell data, 
further reductions in overall computations and thus higher energy efficiency can be achieved.
In this paper, we have demonstrated one such type of driving pattern, namely driving on and off a motorway which is typical during EV use. 
The higher current excitations during these segments provide both low-bias and low-variance parameter estimates 
while reducing the necessary computational cost associated with the TLS algorithm.
The methodology proposed in this paper can be further improved by considering other driving types to be targeted during urban and rural driving. 
Additionally, due to the flexibility of ETREs, other high-level data types can be considered such as pack current, temperature, and SOC.

\section*{DECLARATION OF GENERATIVE AI AND AI-ASSISTED TECHNOLOGIES IN THE WRITING PROCESS}
During the preparation of this work the authors used GitHub Copilot and ChatGPT in the writing of the paper for suggesting text, phrasing, grammatical advice, and basic typesetting.
After using these tools, the authors reviewed and edited the content as needed and take full responsibility for the content of the publication.

\bibliography{references}             

@article{zheng_online_2024,
    title = {Online {Sensorless} {Temperature} {Estimation} of {Lithium}-{Ion} {Batteries} {Through} {Electro}-{Thermal} {Coupling}},
    volume = {29},
    issn = {1941-014X},
    doi = {10.1109/TMECH.2024.3367291},
    number = {6},
    journal = {IEEE/ASME Transactions on Mechatronics},
    author = {Zheng, Yusheng and Che, Yunhong and Hu, Xiaosong and Sui, Xin and Teodorescu, Remus},
    month = dec,
    year = {2024},
    pages = {4156--4167},
}

@inproceedings{lopez_microscopic_2018,
    address = {Maui, HI},
    title = {Microscopic {Traffic} {Simulation} using {SUMO}},
    isbn = {978-1-7281-0321-1 978-1-7281-0323-5},
    doi = {10.1109/ITSC.2018.8569938},
    language = {en},
    booktitle = {2018 21st {International} {Conference} on {Intelligent} {Transportation} {Systems} ({ITSC})},
    publisher = {IEEE},
    author = {Lopez, Pablo Alvarez and Wiessner, Evamarie and Behrisch, Michael and Bieker-Walz, Laura and Erdmann, Jakob and Flotterod, Yun-Pang and Hilbrich, Robert and Lucken, Leonhard and Rummel, Johannes and Wagner, Peter},
    month = nov,
    year = {2018},
    pages = {2575--2582},
}

@inproceedings{weinreich_tracksim_2025,
    address = {Paris, France},
    title = {{TRACKSIM}: {A} {Multi}-{Level} {Simulation} {Framework} for {Near}-{Life} {Battery} {Data} {Generation}},
    copyright = {Creative Commons Attribution 4.0 International},
    shorttitle = {{TRACKSIM}},
    doi = {10.34746/epe2025-0180},
    language = {en},
    urldate = {2025-11-28},
    booktitle = {The 26th {European} {Conference} on {Power} {Electronics} and {Applications}},
    publisher = {GDR SEEDS France \& EPE Association},
    author = {Weinreich, Nicolai A. and Sui, Xin and Teodorescu, Remus and Larsen, Kim G.},
    month = mar,
    year = {2025},
}

@article{zhao_lithium_2024,
    title = {Lithium battery state of charge estimation based on improved variable forgetting factor recursive least squares method and adaptive {Kalman} filter joint algorithm},
    volume = {100},
    issn = {2352152X},
    doi = {10.1016/j.est.2024.113392},
    language = {en},
    journal = {Journal of Energy Storage},
    author = {Zhao, Jinhui and Qian, Xinxin and Jiang, Bing},
    month = oct,
    year = {2024},
    pages = {113392},
}

@article{lao_novel_2018,
    title = {A {Novel} {Method} for {Lithium}-{Ion} {Battery} {Online} {Parameter} {Identification} {Based} on {Variable} {Forgetting} {Factor} {Recursive} {Least} {Squares}},
    volume = {11},
    issn = {1996-1073},
    doi = {10.3390/en11061358},
    language = {en},
    number = {6},
    journal = {Energies},
    author = {Lao, Zizhou and Xia, Bizhong and Wang, Wei and Sun, Wei and Lai, Yongzhi and Wang, Mingwang},
    month = may,
    year = {2018},
    pages = {1358},
}

@article{trongnukul_robust_2025,
    title = {Robust noise-correction recursive least square method for parameter identification of equivalent circuit model in battery management system using {Bayes}’ theorem-based preprocessing technique},
    volume = {107},
    issn = {0948-7921, 1432-0487},
    doi = {10.1007/s00202-024-02371-2},
    language = {en},
    number = {7},
    journal = {Electrical Engineering},
    author = {Trongnukul, Napat and Masomtob, Manop and Fuengwarodsakul, Nisai H.},
    month = jul,
    year = {2025},
    pages = {8531--8547},
}

@article{du_information_2022,
    title = {An {Information} {Appraisal} {Procedure}: {Endows} {Reliable} {Online} {Parameter} {Identification} to {Lithium}-{Ion} {Battery} {Model}},
    volume = {69},
    copyright = {https://ieeexplore.ieee.org/Xplorehelp/downloads/license-information/IEEE.html},
    issn = {0278-0046, 1557-9948},
    shorttitle = {An {Information} {Appraisal} {Procedure}},
    doi = {10.1109/TIE.2021.3091920},
    language = {en},
    number = {6},
    journal = {IEEE Transactions on Industrial Electronics},
    author = {Du, Xinghao and Meng, Jinhao and Zhang, Yingmin and Huang, Xinrong and Wang, Shunliang and Liu, Ping and Liu, Tianqi},
    month = jun,
    year = {2022},
    pages = {5889--5899},
}

@article{cai_novel_2024,
    title = {A novel hybrid electrochemical equivalent circuit model for online battery management systems},
    volume = {99},
    issn = {2352152X},
    doi = {10.1016/j.est.2024.113142},
    language = {en},
    journal = {Journal of Energy Storage},
    author = {Cai, Chengxi and Gong, You and Fotouhi, Abbas and Auger, Daniel J.},
    month = oct,
    year = {2024},
    pages = {113142},
}

@inproceedings{weinreich_data_2025,
    address = {Bengaluru, India},
    title = {A {Data}-{Selection} {Framework} for {Data}-{Efficient} {Battery} {Parameter} {Estimation}},
    copyright = {https://doi.org/10.15223/policy-029},
    isbn = {979-8-3315-1886-8},
    doi = {10.1109/ECCE-Asia63110.2025.11112005},
    language = {en},
    booktitle = {2025 {IEEE} {Energy} {Conversion} {Congress} \& {Exposition} {Asia} ({ECCE}-{Asia})},
    publisher = {IEEE},
    author = {Weinreich, Nicolai A. and Teodorescu, Remus and Larsen, Kim G.},
    month = may,
    year = {2025},
    pages = {1--6},
}

@article{fogelquist_data_2023,
    title = {Data selection framework for battery state of health related parameter estimation under system uncertainties},
    volume = {18},
    issn = {25901168},
    doi = {10.1016/j.etran.2023.100283},
    journal = {eTransportation},
    author = {Fogelquist, Jackson and Lin, Xinfan},
    year = {2023},

}

@article{wei_online_2018,
    title = {Online {Model} {Identification} and {State}-of-{Charge} {Estimate} for {Lithium}-{Ion} {Battery} {With} a {Recursive} {Total} {Least} {Squares}-{Based} {Observer}},
    volume = {65},
    issn = {1557-9948},
    doi = {10.1109/TIE.2017.2736480},
    number = {2},
    journal = {IEEE Transactions on Industrial Electronics},
    author = {Wei, Zhongbao and Zou, Changfu and Leng, Feng and Soong, Boon Hee and Tseng, King-Jet},
    month = feb,
    year = {2018},
    pages = {1336--1346},
}

@article{du_online_2023,
    title = {Online {Identification} of {Lithium}-ion {Battery} {Model} {Parameters} with {Initial} {Value} {Uncertainty} and {Measurement} {Noise}},
    volume = {36},
    issn = {2192-8258},
    doi = {10.1186/s10033-023-00846-0},
    language = {en},
    number = {1},
    journal = {Chinese Journal of Mechanical Engineering},
    author = {Du, Xinghao and Meng, Jinhao and Liu, Kailong and Zhang, Yingmin and Wang, Shunli and Peng, Jichang and Liu, Tianqi},
    month = jan,
    year = {2023},
    pages = {7},
}

@InProceedings{etrepaper,
author="Mu{\~{n}}iz, Marco
and Miku{\v{c}}ionis, Marius
and Larsen, Kim G.",
editor="K{\"o}nighofer, Bettina
and Torfah, Hazem",
title="Extended Timed Regular Expressions",
booktitle="Runtime Verification",
year="2025",
publisher="Springer Nature Switzerland",
address="Cham",
pages="233--251",
isbn="978-3-032-05435-7",
doi="10.1007/978-3-032-05435-7\_14",
}

@inproceedings{AsarinKleeneTheorem,
 author = {Asarin, E. and Caspi, P. and Maler, O.},
 title = {A Kleene theorem for timed automata},
 booktitle = {Proceedings of the 12th Annual IEEE Symposium on Logic in Computer Science},
 series = {LICS '97},
 year = {1997},
 isbn = {0-8186-7925-5},
 pages = {160--},
 acmid = {788856},
 publisher = {IEEE Computer Society},
 address = {Washington, DC, USA},
 doi={10.1109/LICS.1997.614944}
}

@incollection{kleene1956regex,
  author    = {Kleene, Stephen Cole},
  title     = {Representation of Events in Nerve Nets and Finite Automata},
  booktitle = {Automata Studies},
  editor    = {Shannon, Claude E. and McCarthy, John},
  pages     = {3--41},
  publisher = {Princeton University Press},
  year      = {1956},
  address   = {Princeton, NJ},
  isbn      = {0691079161},
  doi       = {10.1515/9781400882618-002}
}
\end{document}